\documentclass[runningheads]{llncs}
\usepackage[misc]{ifsym}
\usepackage{graphicx} 
\usepackage[normalem]{ulem}
\useunder{\uline}{\ul}{}
\usepackage{graphicx} 
\usepackage[colorlinks=true,allcolors=blue]{hyperref}
\usepackage{amsmath,amssymb,amsfonts,bm}

\usepackage{multirow}
\usepackage{booktabs}
\usepackage{placeins}
\usepackage{subfigure} 
\usepackage{floatrow}
\newfloatcommand{capbtabbox}{table}[][\FBwidth]
\begin{document}
\title{Region Attention Transformer for Medical Image Restoration}

\author{
Zhiwen Yang\inst{1} \and
Haowei Chen\inst{1} \and 
Ziniu Qian\inst{1} \and 
Yang Zhou\inst{1} \and
Hui Zhang\inst{2} \and 
Dan Zhao\inst{3} \and 
Bingzheng Wei\inst{4} \and 
Yan Xu\inst{1}$^{(\textrm{\Letter},\thanks{Corresponding author})}$
} 


\authorrunning{Yang et al.}

\institute{
School of Biological Science and Medical Engineering, State Key Laboratory of Software Development Environment, Key Laboratory of Biomechanics and Mechanobiology of Ministry of Education, Beijing Advanced Innovation Center for Biomedical Engineering, Beihang University, Beijing 100191, China
\\
\email{xuyan04@gmail.com} \and 
Department of Biomedical Engineering, Tsinghua University, Beijing 100084, China \and 
Department of Gynecology Oncology, National Cancer Center/National Clinical Research Center for Cancer/Cancer Hospital, Chinese Academy of Medical Sciences and Peking Union Medical College, Beijing 100021, China \and
ByteDance Inc., Beijing 100098, China
}

\maketitle              
\begin{abstract}
Transformer-based methods have demonstrated impressive results in medical image restoration, attributed to the multi-head self-attention (MSA) mechanism in the spatial dimension. However, the majority of existing Transformers conduct attention within fixed and coarsely partitioned regions (\text{e.g.} the entire image or fixed patches), resulting in interference from irrelevant regions and fragmentation of continuous image content. To overcome these challenges, we introduce a novel Region Attention Transformer (RAT) that utilizes a region-based multi-head self-attention mechanism (R-MSA). The R-MSA dynamically partitions the input image into non-overlapping semantic regions using the robust Segment Anything Model (SAM) and then performs self-attention within these regions. This region partitioning is more flexible and interpretable, ensuring that only pixels from similar semantic regions complement each other, thereby eliminating interference from irrelevant regions. Moreover, we introduce a focal region loss to guide our model to adaptively focus on recovering high-difficulty regions. Extensive experiments demonstrate the effectiveness of RAT in various medical image restoration tasks, including PET image synthesis, CT image denoising, and pathological image super-resolution. Code is available at \href{https://github.com/Yaziwel/Region-Attention-Transformer-for-Medical-Image-Restoration.git}{https://github.com/RAT}.

\end{abstract}
\section{Introduction} 

Medical image restoration (MedIR) aims to recover high-quality (HQ) images from their low-quality (LQ) counterparts, encompassing a range of subtasks such as positron emission tomography (PET) image synthesis \cite{zhou2022sgsgan,chen2018dcsrn,chan2018dcnn,zhou2020cyclewagn,luo2022argan,yang2023drmc,jang2023spach}, computed tomography (CT) image denoising \cite{chen2017cnn-ct,yang2018wgan-vgg,chen2017redcnn,wang2023ctformer,liang2020edcnn}, and pathological image super-resolution \cite{li2021Li_pathsr}. It is a challenging problem due to its ill-posed nature, where high-frequency details are missing in the input LQ images. With the advent of deep learning, researchers have explored the use of convolutional neural networks (CNNs) \cite{luo2022argan,liang2020edcnn,li2021Li_pathsr}  and Transformers \cite{yang2023drmc,jang2023spach,wang2023ctformer,liang2021swinir} to develop innovative deep networks for MedIR, significantly advancing this field. 


Recently, Transformers have achieved state-of-the-art performance across a number of MedIR tasks \cite{yang2023drmc,jang2023spach,wang2023ctformer}, attributed to the multi-head self-attention (MSA) mechanism in the spatial dimension. This attention mechanism allows the model to selectively focus on relevant regions within the image, thereby enhancing the MedIR performance \cite{zhang2023lightweight}.  Typically, Transformers conduct attention either across the entire image \cite{vaswani2017attention} or by partitioning it into smaller patches \cite{liu2021swin,liang2021swinir} for efficiency. Despite evidence of their effectiveness in research, there are still two flaws in attention computation. \textbf{(1)} Whether computing attention across the entire image or on patches, it involves roughly aggregating pixels for attention operations. According to the papers \cite{zhao2019sparse_transformer,mei2021non-local,xia2022deformable_attention}, attention potentially assigns high scores to irrelevant pixels, leading to the interaction between pixels from non-similar regions and thereby causing undesirable interference. \textbf{(2)} Dividing the image into smaller patches may result in splitting continuous regions into different patches, thereby disrupting the continuity of the image content and impeding mutual complementarity between similar regions.


In order to address these issues, we propose a novel region-based multi-head self-attention (R-MSA) method that dynamically partitions the input image into fine-grained semantic regions and then performs attention within these semantic regions. This fine-grained partitioning is facilitated by the recently proposed segment anything model (SAM) \cite{kirillov2023SAM}, which exhibits strong segmentation capabilities even in LQ images with severe degradation. Compared to previous attention methods (e.g., \cite{vaswani2017attention,liu2021swin}), which typically treat the entire image as a single region or divide the image into patch-based regions with fixed shapes, our proposed R-MSA offers better flexibility and interpretability. Firstly, through the dynamic semantic region partitioning by SAM, pixels within each region of R-MSA possess similar semantic information \cite{wang2018sft}, which can complement each other, thereby reducing interference from unrelated regions during attention operations and addressing the \textbf{flaw (1)}. Secondly, the semantic information within each region of R-MSA remains complete and continuous, avoiding the disruption of semantic content continuity caused by the coarse patch division in \textbf{flaw (2)}.

In this paper, we introduce a region attention transformer (RAT) for medical image restoration. RAT utilizes the powerful SAM model to partition the input into non-overlapping semantic regions in a data-dependent manner. Subsequently, a novel region-based multi-head self-attention (R-MSA) mechanism is introduced, which conducts attention computation within these semantic regions. Due to its selective attention mechanism, R-MSA is interpretable and ensures that only pixels within the same semantic region complement each other, thereby eliminating interference from irrelevant regions. In addition, considering the varying difficulty levels across distinct image regions, we propose a focal region loss that prioritizes the recovery of high-difficulty regions during training. This is achieved by dynamically assigning loss weights to different regions, with higher weights specifically allocated to the high-difficulty regions. Extensive experiments demonstrate that our proposed RAT achieves state-of-the-art performance in PET image synthesis, CT image denoising, and pathological image super-resolution. The contributions of our work can be summarized as follows: 

\begin{itemize} 
\item We introduce a novel region attention transformer (RAT) for medical image restoration, which incorporates semantic knowledge obtained from the powerful segment anything model (SAM), resulting in a more interpretable framework. RAT could be one of the first methods to utilize SAM to boost medical image restoration.
\item We propose a novel region-based multi-head self-attention mechanism (R-MSA), designed to perform self-attention within semantic regions partitioned by SAM. The R-MSA ensures that attention operations are confined to similar areas, effectively eliminating interference from irrelevant regions.
\item We introduce a focal region loss to prioritize the recovery of high-difficulty regions during training, significantly enhancing image restoration performance.

\end{itemize}

\section{Method} 

\begin{figure*}[t]
	\centering
	\includegraphics[width=\textwidth]{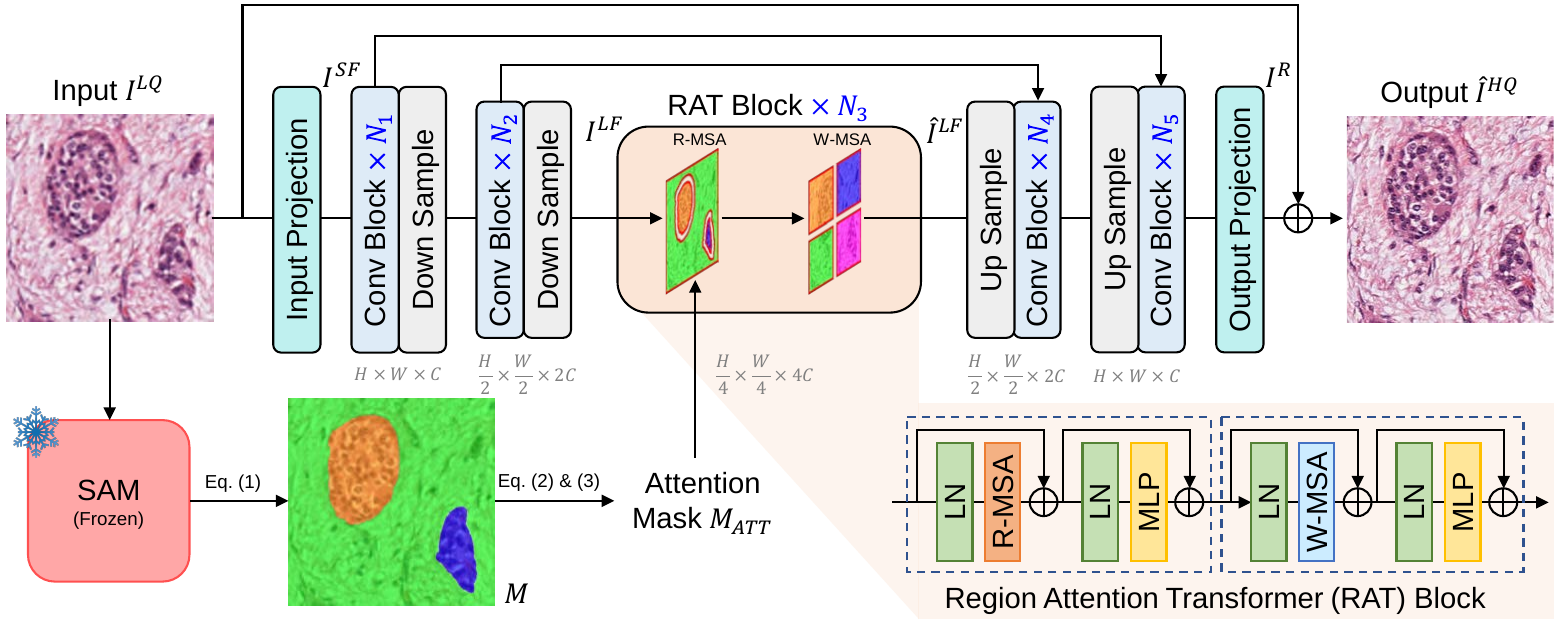}
    \caption{Overall architecture of the proposed region attention transformer (RAT).}
	\label{fig:overall_arch}
\end{figure*}

As is shown in Fig.~\ref{fig:overall_arch}, the proposed region attention transformer (RAT) consists of a U-shaped restoration branch and a segment anything model (SAM) branch that provides region guidance for the restoration process. Specifically, given a low-quality (LQ) medical image $I^{LQ}$, RAT first applies a 3$\times$3 convolution as input projection to obtain shallow features $I^{SF}$. Next, these shallow features $I^{SF}$ undergo a 2-level encoding, resulting in low-resolution latent features $I^{LF}$, with each encoding level consisting of multiple convolution blocks. The latent features $I^{LF}$ are then refined by RAT blocks for modeling long-range dependencies with attention. In this process, the SAM branch provides region guidance to guarantee that the RAT block conducts attention within the same semantic regions and mitigates non-similar content interference. The refined latent features $\hat{I}^{LF}$ then undergo a 2-level decoding and progressively recover the high-resolution features. Finally, a 3$\times$3 convolution is applied as output projection to generate residual image $I^{R}$. The restored output image is obtained by element-wise sum: $\hat{I}^{HQ} = I^{LQ} + I^{R}$. Notably, NAFBlock \cite{chen2022NAFNet} is chosen for convolution blocks in the encoding and decoding stages due to its simplicity and effectiveness. Feature downsampling and upsampling operations are achieved using pixel-unshuffle and pixel-shuffle operations. To assist the restoration process, the encoder features are summed with the decoder features via skip connections. Additionally, a focal region loss is introduced to prioritize the recovery of high-difficulty regions during training. In the following subsection, we will present our proposed RAT block as well as the focal region loss in detail. 

\subsection{Region Attention Transformer Block} 
As illustrated in Fig.~\ref{fig:overall_arch}, the basic layers of the RAT block exhibit a structure similar to the original transformer \cite{vaswani2017attention}, including consecutive modules of MSA and multi-layer perceptron (MLP), along with layer normalization (LN) applied before them. The main distinction lies in our proposed region-based MSA (R-MSA). For a comprehensive understanding, we delve into the RAT block across three parts: region partitioning, region attention, and cross-region connection.

\textbf{Region Partitioning.} Region partitioning determines where attention is conducted. Previous attention methods relying on fixed region partitioning often suffer from interference caused by dissimilar content and disruption of semantic content continuity. To overcome these challenges, we propose to partition the input feature into semantic regions that preserve structural continuity while minimizing interference from irrelevant regions. This is based on a powerful large-scale foundational model: the Segment Anything model (SAM) \cite{kirillov2023SAM}. SAM can accurately segment any object in any image, including medical images, without the need for additional training. Given the input image $I^{LQ}$, the region partitioning masks through SAM can be expressed as:
\begin{equation}
M=\operatorname{Postprocess}(\operatorname{SAM}(I^{LQ})),
\end{equation}
where $\operatorname{Postprocess}(\cdot)$ ensures that partitioning bianry masks $M \in \mathbb{R}^{H\times W \times L}$ divide the image into $L$ (which depends on the input image) non-overlapping regions, guaranteeing that each pixel in the image is assigned to a unique region. The pixel belonging to multiple mask regions is categorized into the smallest and finest among them, while the one not belonging to any masks generated by SAM is classified as the background region.

\textbf{Region Attention.} After obtaining the region partitioning binary masks $M$, one straightforward way to compute attention is to iterate through each region and perform attention within it. However, such loop-based computations are highly time-consuming. To address this, we draw inspiration from the Swin Transformer \cite{liu2021swin} to apply masks on attention maps. This allows for effective control over attention regions while leveraging the efficiency of rapid batch computation. Unlike the Swin Transformer, our masks are data-dependent and offer better interpretability. The attention mask can be acquired as follows:

\begin{equation}
M'=\operatorname{Reshape}(\operatorname{NearestInterpolation}(M)),
\label{eq2}
\end{equation}

\begin{equation}
M_{ATT}=\lambda \sum_{i=1}^L\left(1-M_i^{'} M^{'  \top}_i\right), 
\label{eq3}
\end{equation}
where Eq.~\ref{eq2} interpolates $M$ to the target spatial resolution $H' \times W'$ and reshapes the result to $M' \in \mathbb{R}^{H'W' \times L}$. Eq.~\ref{eq3} acquires the attention mask $M_{ATT}  \in \mathbb{R}^{H'W' \times H'W'}$ by assigning negative infinite values, denoted by 
$\lambda$, to the correlation scores between pixels from different semantic regions. Consequently, given the $Q$, $K$, $V \in \mathbb{R}^{H'W' \times C'}$, which are respectively the query, key, and value from the linear projection of input feature 
$X \in \mathbb{R}^{H'W' \times C'}$, the mechanism of R-MSA can be formulated as follows:
\begin{equation}
\operatorname{R-MSA}(Q, K, V)=\operatorname{Softmax}\left(M_{ATT} + Q K^T/\sqrt{d} \right) V,
\end{equation} 
where $d$ is the attention head number. The attention mask $M_{ATT}$ is added to exclude potential correlations between pixels from different semantic regions, thus constraining the attention range within individual semantic regions.

\textbf{Cross-Region Connection.} To maintain interaction between semantic regions, the RAT block utilizes two consecutive transformer layers, with a window-based MSA (W-MSA) \cite{liu2021swin} transformer layer immediately following the R-MSA transformer layer, as illustrated in Fig.~\ref{fig:overall_arch}. The W-MSA transformer layer partitions the input feature into non-overlapping windows, some of which cover the boundaries between different semantic regions in R-MSA. Performing attention computation within these windows facilitates connections between different regions in R-MSA. 

\subsection{Focal Region Loss} 
Considering that the restoration difficulty varies across different regions of the image, directing the model's focus towards the high-difficulty region aids in learning the recovery of intricate details within the image. Hence, we follow focal loss \cite{lin2017focalloss} and introduce a focal region loss utilizing partitioning masks $M$ to prioritize the recovery of high-difficulty regions: 

\begin{equation}
L_{FR}=\sum_{i=1}^L (1+\gamma w_i) \left|\hat{I}_{M_i}^{H Q}-I_{M_i}^{H Q}\right|, 
\end{equation}
where $\hat{I}_{M_i}^{H Q}$ and $I_{M_i}^{H Q}$ denote the $M_i$ region of the restored image and high-quality image, respectively. The normalized weighting parameter $w_i =\frac{\left|\hat{I}_{M_i}^{H Q}-I_{M_i}^{H Q}\right|^{\delta}}{\max \left\{\left|\hat{I}_{M j}^{H Q}-I_{M_j}^{H Q}\right|^{\delta}\right\}_{j=1}^L}$ dynamically assigns weights to prioritize high-difficulty regions. $\delta$ and $\gamma$ are scaling factors for flexibility. $L_{FR}$ degenerates to $L_1$ loss when $\gamma=0$.

\section{Experiments and Results} 
\subsection{Dataset} 
We conduct experiments on three typical medical image restoration tasks: PET image synthesis, CT image denoising, and pathological image super-resolution. 

\textbf{PET Image Synthesis.} We acquire 115 HQ PET images using the PolarStar m660 PET/CT system in list mode, with an injection dose of 293MBq $^{18}$F-FDG. LQ PET images are generated through list mode decimation with a dose reduction factor of 12 \cite{xiang2017auto-contextcnn}. Both HQ and LQ PET images are reconstructed using the standard OSEM method \cite{hudson1994osem}. Each PET image has 3D shapes of 192$\times$192$\times$416 and is divided into 192 2D slices sized 192$\times$416. Slices containing only air are excluded. Patient data are divided into 90 for training and 25 for testing. 

\begin{table}[t] 
\centering
\caption{Comparisons of PET image synthesis on the private dataset.} 
\resizebox{\textwidth}{!}{
\begin{tabular}{c|c|c|c|c|c|c}
\toprule
Method & DCSRN \cite{chen2018dcsrn,zhou2022sgsgan}            & Xiang's \cite{xiang2017auto-contextcnn} & DCNN \cite{chan2018dcnn}             & CycleWGAN \cite{zhou2020cyclewagn}        & AR-GAN \cite{luo2022argan}           & \textbf{RAT}              \\ \midrule
PSNR↑  & 40.3702 ± 2.5205 & 40.4753 ± 2.4685 & 40.6273 ± 2.4153 & 40.6238 ± 2.5270 & 40.5832 ± 2.6436 & \textbf{40.9487 ± 2.5233} \\ \hline
SSIM↑  & 0.9686 ± 0.0121  & 0.9688 ± 0.0121  & 0.9699 ± 0.0116  & 0.9700 ± 0.0118  & 0.9702 ± 0.0119  & \textbf{0.9712 ± 0.0113}  \\ \bottomrule
\end{tabular}
}
\label{tab:pet}
\end{table}  

\begin{figure*}[t]
	\centering
	\includegraphics[width=\textwidth]{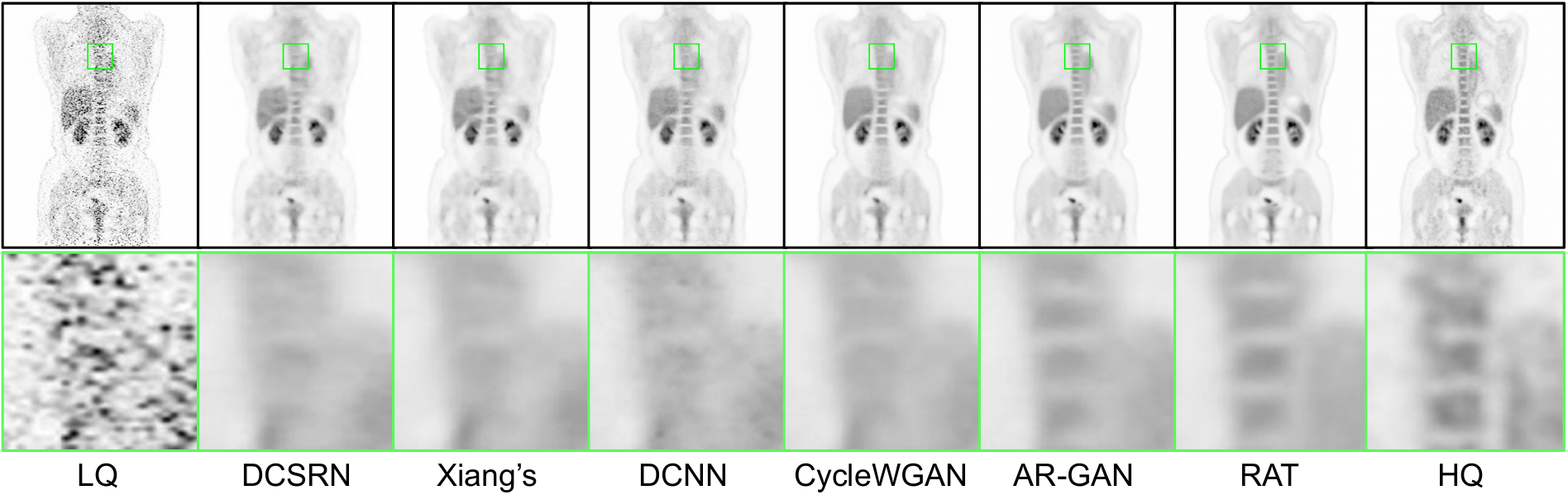}
    \caption{Visual comparison among methods on PET image synthesis. }
	\label{fig:pet}
\end{figure*}

\textbf{CT Image Denoising.} We employ the dataset from the 2016 NIH AAPM-Mayo Clinic Low-Dose CT Grand Challenge \cite{mccollough2017AAPM}, which comprises paired HQ CT images taken with normal dose and LQ CT images with quarter dose, each sized 512x512. These images are collected from 10 patients, with a division of 9 for training and 1 for testing. 

\textbf{Pathological Image Super-Resolution.} We utilize the TMA dataset \cite{drifka2016TMA}, which comprises 573 HQ pathological images with an average size of 3249$\times$3249. The LQ images are acquired via 4$\times$ bicubic downsampling and subsequently upsampled back to the original resolution. We partition these images into 460 for training and 113 for testing.

\subsection{Implementation} 
The convolution blocks at different levels of the model are set to $N_1=N_2=N_4=N_5=2$, with $N_3=12$ for RAT blocks. The number of channels after input projection is set to $C=64$. Both R-MSA and W-MSA have 8 attention heads. The value of negative infinity in the R-MSA attention mask $M_{ATT}$ is set to $\lambda=-1000$. The window size of W-MSA is set to 4. The well-trained SAM model \cite{kirillov2023SAM} with the ViT-B backbone is used for region partitioning. The scaling factors in the loss function $L_{FR}$ are set to $\gamma=1e^{-3}$ and $\delta=1$. The patch size for training CT image denoising and pathological image super-resolution is set to 256$\times$256, while for PET image synthesis, it is set to 192$\times$192. The model is trained using the Adam optimizer for 200K iterations, starting with an initial learning rate of $2e^{-4}$, gradually reduced to $1e^{-6}$ using cosine annealing.

\begin{table}[t] 
\centering
\caption{Comparisons of CT image denoising on the AAPM \cite{mccollough2017AAPM} dataset.} 
\resizebox{\textwidth}{!}{
\begin{tabular}{c|c|c|c|c|c|c}
\toprule
Method & CNN \cite{chan2018dcnn}              & WGAN-VGG \cite{yang2018wgan-vgg}         & REDCNN \cite{chen2017redcnn}           & CTformer \cite{wang2023ctformer}         & EDCNN \cite{liang2020edcnn}            & \textbf{RAT}              \\ \midrule
PSNR↑  & 41.9620 ± 0.7087 & 43.6399 ± 1.4728 & 45.5887 ± 1.3911 & 45.6322 ± 1.4139 & 45.6986 ± 1.3931 & \textbf{45.8933 ± 1.4230} \\ \hline
SSIM↑  & 0.9627 ± 0.0108  & 0.9610 ± 0.0156  & 0.9742 ± 0.0100  & 0.9743 ± 0.0101  & 0.9746 ± 0.0098  & \textbf{0.9754 ± 0.0098}  \\ \bottomrule
\end{tabular}
}
\label{tab:ct}
\end{table}  

\begin{figure*}[t]
	\centering
	\includegraphics[width=\textwidth]{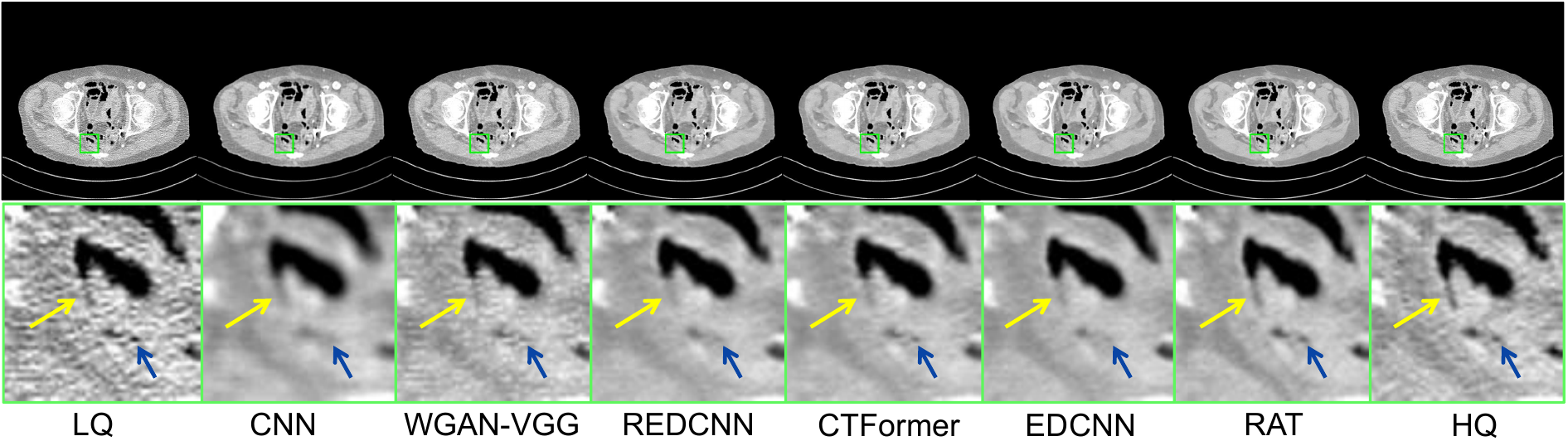}
    \caption{Visual comparison among methods on CT image denoising.}
	\label{fig:ct}
\end{figure*}

\subsection{Comparison Experiment} 
We assess RAT against five PET image synthesis methods, five CT image denoising methods, and five pathological image super-resolution methods. Specifically, the PET image synthesis baselines consist of DCSRN \cite{chen2018dcsrn,zhou2022sgsgan}, Xiang's method \cite{xiang2017auto-contextcnn}, DCNN \cite{chan2018dcnn}, CycleWGAN \cite{zhou2020cyclewagn}, and AR-GAN \cite{luo2022argan}. The CT image denoising baselines encompass CNN \cite{chen2017cnn-ct}, WGAN-VGG \cite{yang2018wgan-vgg}, REDCNN \cite{chen2017redcnn}, EDCNN \cite{liang2020edcnn}, and CTformer \cite{wang2023ctformer}. The pathological image super-resolution methods include SRCNN \cite{dong2015srcnn}, EDSR \cite{lim2017edsr}, RCAN \cite{zhang2018rcan}, Li's method \cite{li2021Li_pathsr}, and SwinIR \cite{liang2021swinir}. To evaluate these methods, we employ commonly used metrics including PSNR and SSIM.

\begin{figure*}[t]
	\centering
	\includegraphics[width=\textwidth]{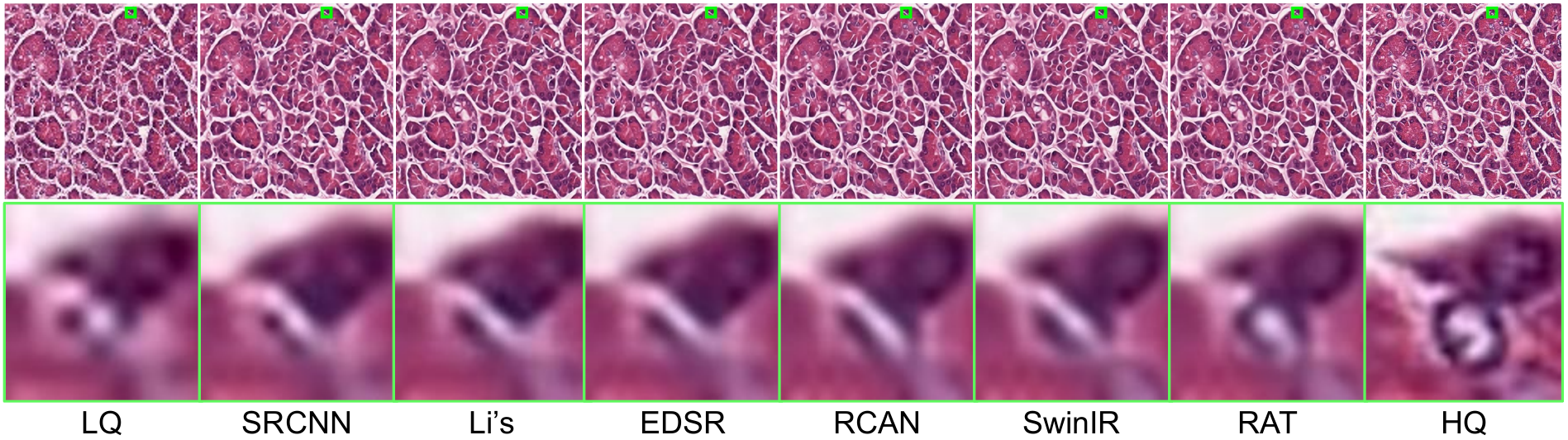}
    \caption{Visual comparison among methods on pathological image super-resolution.}
	\label{fig:pathoology}
\end{figure*}

The quantitative comparison results for PET image synthesis, CT image denoising, and pathological image super-resolution are presented in Tables \ref{tab:pet}, \ref{tab:ct}, and \ref{tab:pathology}, respectively. It is evident that our proposed RAT surpasses all methods in comparison, achieving state-of-the-art performance across all three tasks. In contrast to CNN-based methods, RAT demonstrates superior performance by leveraging long-range dependencies of attention. Compared to Transformer-based methods such as CTformer \cite{wang2023ctformer} and SwinIR \cite{liang2021swinir}, RAT exhibits superior results attributed to the flexible and accurate region partitioning of R-MSA, which effectively eliminates potential interferences from irrelevant regions and preserves semantic content continuity. Visual comparisons across the three tasks are depicted in Figs. \ref{fig:pet}, \ref{fig:ct}, and \ref{fig:pathoology}, respectively, clearly illustrating that our proposed RAT excels in recovering structures and details.

\begin{table}[t] 
\centering
\caption{Comparisons of pathological image super-resolution on the TMA \cite{drifka2016TMA} dataset.} 
\resizebox{\textwidth}{!}{
\begin{tabular}{c|c|c|c|c|c|c}
\toprule
Method & SRCNN \cite{dong2015srcnn}            & Li's \cite{li2021Li_pathsr}            & EDSR \cite{lim2017edsr}             & RCAN \cite{zhang2018rcan}            & SwinIR \cite{liang2021swinir}           & \textbf{RAT}              \\ \midrule
PSNR↑  & 22.5851 ± 1.1663 & 22.9006 ± 1.1671 & 22.9904 ± 1.1697 & 23.2188 ± 1.1742 & 23.2477 ± 1.1824 & \textbf{23.3434 ± 1.1826} \\ \hline
SSIM↑  & 0.7071 ± 0.0579  & 0.7204 ± 0.0588  & 0.7276 ± 0.0574  & 0.7385 ± 0.0573  & 0.7398 ± 0.0576  & \textbf{0.7436 ± 0.0573}  \\ \bottomrule
\end{tabular}
}
\label{tab:pathology}
\end{table}  

\subsection{Ablation Study} 
We conduct ablation experiments on the attention mechanism and loss components proposed in this paper. Regarding the attention mechanism, we compare our proposed R-MSA with other attention mechanisms such as standard MSA \cite{vaswani2017attention} and W-MSA \cite{liu2021swin} by replacing R-MSA with these attention mechanisms. The experimental results, as shown in Table.~\ref{tab:ablation}, demonstrate that our proposed R-MSA significantly outperforms MSA and W-MSA across the three tasks. This is attributed to the finer region partitioning, which can more effectively avoid interference from irrelevant regions, thereby achieving better image restoration. Furthermore, Table.~\ref{tab:ablation} also indicates a stable improvement in the proposed focal region loss across the three different tasks, demonstrating its effectiveness.

\begin{table}[h] 
\centering
\caption{Ablation experiments on components of attention mechanism and loss.} 
\resizebox{\textwidth}{!}{
\begin{tabular}{c|c|cc|cc|cc}
\toprule
\multirow{2}{*}{Component} & \multirow{2}{*}{Method} & \multicolumn{2}{c|}{PET Image Synthesis}                                  & \multicolumn{2}{c|}{CT Image Denoising}                                         & \multicolumn{2}{c}{Pathological Image Super-Resolution}                   \\ \cline{3-8} 
                           &                         & \multicolumn{1}{c|}{PSNR↑}                     & SSIM↑                    & \multicolumn{1}{c|}{PSNR↑}                     & SSIM↑                    & \multicolumn{1}{c|}{PSNR↑}                     & SSIM↑                    \\ \midrule
\multirow{3}{*}{Attention} & MSA                     & \multicolumn{1}{c|}{40.5324 ± 2.2975}          & 0.9704 ± 0.0115          & \multicolumn{1}{c|}{45.8102 ± 1.4170}          & 0.9749 ± 0.0098          & \multicolumn{1}{c|}{23.2581 ± 1.1784}          & 0.7406 ± 0.0576          \\
                           & W-MSA                   & \multicolumn{1}{c|}{40.6983 ± 2.4798}          & 0.9702 ± 0.0116          & \multicolumn{1}{c|}{45.8616 ± 1.4191}          & 0.9753 ± 0.0098          & \multicolumn{1}{c|}{23.3035 ± 1.1847}          & 0.7423 ± 0.0572          \\
                           & R-MSA                   & \multicolumn{1}{c|}{\textbf{40.9487 ± 2.5233}} & \textbf{0.9712 ± 0.0113} & \multicolumn{1}{c|}{\textbf{45.8933 ± 1.4230}} & \textbf{0.9754 ± 0.0098} & \multicolumn{1}{c|}{\textbf{23.3434 ± 1.1826}} & \textbf{0.7436 ± 0.0573} \\ \hline
\multirow{2}{*}{Loss}      & $L_1$                   & \multicolumn{1}{c|}{40.8774 ± 2.5206}          & 0.9710  ± 0.0114         & \multicolumn{1}{c|}{45.8810 ± 1.4214}          & 0.9753 ± 0.0098          & \multicolumn{1}{c|}{23.3145 ± 1.1838}          & 0.7427 ± 0.0574          \\
                           & $L_{FR}$                & \multicolumn{1}{c|}{\textbf{40.9487 ± 2.5233}} & \textbf{0.9712 ± 0.0113} & \multicolumn{1}{c|}{\textbf{45.8933 ± 1.4230}} & \textbf{0.9754 ± 0.0098} & \multicolumn{1}{c|}{\textbf{23.3434 ± 1.1826}} & \textbf{0.7436 ± 0.0573} \\ \bottomrule
\end{tabular}
}
\label{tab:ablation}
\end{table}

\subsection{Conclusion} 
In this paper, we introduce a novel region attention transformer (RAT) for medical image restoration. RAT conducts attention within similar semantic regions, facilitating pixels with similar semantic information to complement each other and thereby mitigating interference from non-similar content. Additionally, a focal region loss is introduced to direct the model's focus towards recovering challenging regions. Experiments demonstrate that RAT achieves state-of-the-art performance in PET image synthesis, CT denoising, and pathological image super-resolution.

\begin{credits}
\subsubsection{\ackname} 
This work is supported by the National Natural Science Foundation in China under Grant 62371016, U23B2063, 62022010, and 62176267, the Bejing Natural Science Foundation Haidian District Joint Fund in China under Grant L222032, the Beijing hope run special fund of cancer foundation of China under Grant LC2018L02, the Fundamental Research Funds for the Central University of China from the State Key Laboratory of Software Development Environment in Beihang University in China, the 111 Proiect in China under Grant B13003, the SinoUnion Healthcare Inc. under the eHealth program, the high performance computing (HPC) resources at Beihang University.

\subsubsection{\discintname}
We have no conflicts of interest to disclose.
\end{credits}


%
%
%
%
\bibliographystyle{splncs04}
\bibliography{Paper-0515}

\begin{thebibliography}{10}
\providecommand{\url}[1]{\texttt{#1}}
\providecommand{\urlprefix}{URL }
\providecommand{\doi}[1]{https://doi.org/#1}

\bibitem{zhou2022sgsgan}
Zhou, Y., Yang, Z., Zhang, H., Eric, I., Chang, C., Fan, Y., Xu, Y.: 3d segmentation guided style-based generative adversarial networks for pet synthesis. IEEE Transactions on Medical Imaging  \textbf{41}(8),  2092--2104 (2022)

\bibitem{chen2018dcsrn}
Chen, Y., Xie, Y., Zhou, Z., Shi, F., Christodoulou, A.G., Li, D.: Brain mri super resolution using 3d deep densely connected neural networks. In: 2018 IEEE 15th international symposium on biomedical imaging (ISBI 2018). pp. 739--742. IEEE (2018)

\bibitem{chan2018dcnn}
Chan, C., Zhou, J., Yang, L., Qi, W., Kolthammer, J., Asma, E.: Noise adaptive deep convolutional neural network for whole-body pet denoising. In: 2018 IEEE Nuclear Science Symposium and Medical Imaging Conference Proceedings (NSS/MIC). pp.~1--4. IEEE (2018)

\bibitem{zhou2020cyclewagn}
Zhou, L., Schaefferkoetter, J.D., Tham, I.W., Huang, G., Yan, J.: Supervised learning with cyclegan for low-dose fdg pet image denoising. Medical image analysis  \textbf{65},  101770 (2020)

\bibitem{luo2022argan}
Luo, Y., Zhou, L., Zhan, B., Fei, Y., Zhou, J., Wang, Y., Shen, D.: Adaptive rectification based adversarial network with spectrum constraint for high-quality pet image synthesis. Medical Image Analysis  \textbf{77},  102335 (2022)

\bibitem{yang2023drmc}
Yang, Z., Zhou, Y., Zhang, H., Wei, B., Fan, Y., Xu, Y.: Drmc: A generalist model with dynamic routing for multi-center pet image synthesis. In: International Conference on Medical Image Computing and Computer-Assisted Intervention. pp. 36--46. Springer (2023)

\bibitem{jang2023spach}
Jang, S.I., Pan, T., Li, Y., Heidari, P., Chen, J., Li, Q., Gong, K.: Spach transformer: spatial and channel-wise transformer based on local and global self-attentions for pet image denoising. IEEE transactions on medical imaging  (2023)

\bibitem{chen2017cnn-ct}
Chen, H., Zhang, Y., Zhang, W., Liao, P., Li, K., Zhou, J., Wang, G.: Low-dose ct denoising with convolutional neural network. In: 2017 IEEE 14th International Symposium on Biomedical Imaging (ISBI 2017). pp. 143--146. IEEE (2017)

\bibitem{yang2018wgan-vgg}
Yang, Q., Yan, P., Zhang, Y., Yu, H., Shi, Y., Mou, X., Kalra, M.K., Zhang, Y., Sun, L., Wang, G.: Low-dose ct image denoising using a generative adversarial network with wasserstein distance and perceptual loss. IEEE transactions on medical imaging  \textbf{37}(6),  1348--1357 (2018)

\bibitem{chen2017redcnn}
Chen, H., Zhang, Y., Kalra, M.K., Lin, F., Chen, Y., Liao, P., Zhou, J., Wang, G.: Low-dose ct with a residual encoder-decoder convolutional neural network. IEEE transactions on medical imaging  \textbf{36}(12),  2524--2535 (2017)

\bibitem{wang2023ctformer}
Wang, D., Fan, F., Wu, Z., Liu, R., Wang, F., Yu, H.: Ctformer: convolution-free token2token dilated vision transformer for low-dose ct denoising. Physics in Medicine \& Biology  \textbf{68}(6),  065012 (2023)

\bibitem{liang2020edcnn}
Liang, T., Jin, Y., Li, Y., Wang, T.: Edcnn: Edge enhancement-based densely connected network with compound loss for low-dose ct denoising. In: 2020 15th IEEE International Conference on Signal Processing (ICSP). vol.~1, pp. 193--198. IEEE (2020)

\bibitem{li2021Li_pathsr}
Li, B., Keikhosravi, A., Loeffler, A.G., Eliceiri, K.W.: Single image super-resolution for whole slide image using convolutional neural networks and self-supervised color normalization. Medical Image Analysis  \textbf{68},  101938 (2021)

\bibitem{liang2021swinir}
Liang, J., Cao, J., Sun, G., Zhang, K., Van~Gool, L., Timofte, R.: Swinir: Image restoration using swin transformer. In: Proceedings of the IEEE/CVF international conference on computer vision. pp. 1833--1844 (2021)

\bibitem{zhang2023lightweight}
Zhang, A., Ren, W., Liu, Y., Cao, X.: Lightweight image super-resolution with superpixel token interaction. In: Proceedings of the IEEE/CVF International Conference on Computer Vision. pp. 12728--12737 (2023)

\bibitem{vaswani2017attention}
Vaswani, A., Shazeer, N., Parmar, N., Uszkoreit, J., Jones, L., Gomez, A.N., Kaiser, {\L}., Polosukhin, I.: Attention is all you need. Advances in neural information processing systems  \textbf{30} (2017)

\bibitem{liu2021swin}
Liu, Z., Lin, Y., Cao, Y., Hu, H., Wei, Y., Zhang, Z., Lin, S., Guo, B.: Swin transformer: Hierarchical vision transformer using shifted windows. In: Proceedings of the IEEE/CVF international conference on computer vision. pp. 10012--10022 (2021)

\bibitem{zhao2019sparse_transformer}
Zhao, G., Lin, J., Zhang, Z., Ren, X., Su, Q., Sun, X.: Explicit sparse transformer: Concentrated attention through explicit selection. arXiv preprint arXiv:1912.11637  (2019)

\bibitem{mei2021non-local}
Mei, Y., Fan, Y., Zhou, Y.: Image super-resolution with non-local sparse attention. In: Proceedings of the IEEE/CVF conference on computer vision and pattern recognition. pp. 3517--3526 (2021)

\bibitem{xia2022deformable_attention}
Xia, Z., Pan, X., Song, S., Li, L.E., Huang, G.: Vision transformer with deformable attention. In: Proceedings of the IEEE/CVF conference on computer vision and pattern recognition. pp. 4794--4803 (2022)

\bibitem{kirillov2023SAM}
Kirillov, A., Mintun, E., Ravi, N., Mao, H., Rolland, C., Gustafson, L., Xiao, T., Whitehead, S., Berg, A.C., Lo, W.Y., et~al.: Segment anything. arXiv preprint arXiv:2304.02643  (2023)

\bibitem{wang2018sft}
Wang, X., Yu, K., Dong, C., Loy, C.C.: Recovering realistic texture in image super-resolution by deep spatial feature transform. In: Proceedings of the IEEE conference on computer vision and pattern recognition. pp. 606--615 (2018)

\bibitem{chen2022NAFNet}
Chen, L., Chu, X., Zhang, X., Sun, J.: Simple baselines for image restoration. In: European Conference on Computer Vision. pp. 17--33. Springer (2022)

\bibitem{lin2017focalloss}
Lin, T.Y., Goyal, P., Girshick, R., He, K., Doll{\'a}r, P.: Focal loss for dense object detection. In: Proceedings of the IEEE international conference on computer vision. pp. 2980--2988 (2017)

\bibitem{xiang2017auto-contextcnn}
Xiang, L., Qiao, Y., Nie, D., An, L., Lin, W., Wang, Q., Shen, D.: Deep auto-context convolutional neural networks for standard-dose pet image estimation from low-dose pet/mri. Neurocomputing  \textbf{267},  406--416 (2017)

\bibitem{hudson1994osem}
Hudson, H.M., Larkin, R.S.: Accelerated image reconstruction using ordered subsets of projection data. IEEE transactions on medical imaging  \textbf{13}(4),  601--609 (1994)

\bibitem{mccollough2017AAPM}
McCollough, C.H., Bartley, A.C., Carter, R.E., Chen, B., Drees, T.A., Edwards, P., Holmes~III, D.R., Huang, A.E., Khan, F., Leng, S., et~al.: Low-dose ct for the detection and classification of metastatic liver lesions: results of the 2016 low dose ct grand challenge. Medical physics  \textbf{44}(10),  e339--e352 (2017)

\bibitem{drifka2016TMA}
Drifka, C.R., Loeffler, A.G., Mathewson, K., Keikhosravi, A., Eickhoff, J.C., Liu, Y., Weber, S.M., Kao, W.J., Eliceiri, K.W.: Highly aligned stromal collagen is a negative prognostic factor following pancreatic ductal adenocarcinoma resection. Oncotarget  \textbf{7}(46),  76197 (2016)

\bibitem{dong2015srcnn}
Dong, C., Loy, C.C., He, K., Tang, X.: Image super-resolution using deep convolutional networks. IEEE transactions on pattern analysis and machine intelligence  \textbf{38}(2),  295--307 (2015)

\bibitem{lim2017edsr}
Lim, B., Son, S., Kim, H., Nah, S., Mu~Lee, K.: Enhanced deep residual networks for single image super-resolution. In: Proceedings of the IEEE conference on computer vision and pattern recognition workshops. pp. 136--144 (2017)

\bibitem{zhang2018rcan}
Zhang, Y., Li, K., Li, K., Wang, L., Zhong, B., Fu, Y.: Image super-resolution using very deep residual channel attention networks. In: Proceedings of the European conference on computer vision (ECCV). pp. 286--301 (2018)

\end{thebibliography}
\end{document}